\documentclass[a4paper,fleqn]{cas-dc}

\usepackage{graphicx}
\usepackage[font=scriptsize, justification=centering]{caption}
\usepackage{subcaption}
\usepackage[switch]{lineno}
\usepackage{multirow}
\usepackage{enumitem}
\usepackage{listings}
\usepackage{xspace}
\usepackage[flushleft]{threeparttable}
\usepackage[numbers]{natbib}

\newcommand{\hl}[1]{#1}

\def\tsc#1{\csdef{#1}{\textsc{\lowercase{#1}}\xspace}}
\tsc{WGM}
\tsc{QE}
\tsc{EP}
\tsc{PMS}
\tsc{BEC}
\tsc{DE}

\newdefinition{definition}{Definition}
\newcommand{\etal}{\textit{et al.}}
\newcommand{\vt}{vulnerability type }

\newcommand{\rp}{$Rule^+$}
\newcommand{\rn}{$Rule^-$}
\newcommand{\base}{\textsc{Base}\xspace}
\newcommand{\ebase}{\textsc{Enhanced Base}\xspace}

\definecolor{ao}{rgb}{0.0, 0.5, 0.0}
\begin{document}
\let\WriteBookmarks\relax
\def\floatpagepagefraction{1}
\def\textpagefraction{.001}

\author{Hieu Dinh Vo}[orcid=0000-0002-9407-1971]

\ead{hieuvd@vnu.edu.vn}

\affiliation{organization={Faculty of Information Technology, VNU University of Engineering and Technology},
    city={Hanoi},
    country={Vietnam}}

\author{Son Nguyen}[orcid=0000-0002-8970-9870]
\ead{sonnguyen@vnu.edu.vn}
\cormark[1]

\cortext[cor1]{Corresponding author}
\shorttitle{Can An Old Fashioned Feature Extraction and A Light-weight Model Improve VIT Performance?}
\shortauthors{Vo et~al.}
\title [mode = title]{Can An Old Fashioned Feature Extraction and A Light-weight Model Improve Vulnerability Type Identification Performance?}

\begin{abstract}
Recent advances in automated vulnerability detection have achieved potential results in helping developers determine vulnerable components. However, after detecting vulnerabilities, investigating to fix vulnerable code is a non-trivial task. In fact, the types of vulnerability, such as \textit{buffer overflow} or \textit{memory corruption}, could help developers quickly understand the nature of the weaknesses and localize vulnerabilities for security analysis.
In this work, we investigate the problem of vulnerability type identification (VTI). The problem is modeled as the multi-label classification task, which could be effectively addressed by ``\textit{pre-training, then fine-tuning}'' framework with deep pre-trained embedding models. 
We evaluate the performance of the well-known and advanced pre-trained models for VTI on a large set of vulnerabilities. Surprisingly, their performance is not much better than that of the classical baseline approach with an old-fashioned bag-of-word, TF-IDF. Meanwhile, these deep neural network approaches cost much more resources and require GPU. We also introduce a lightweight independent component to refine the predictions of the baseline approach. Our idea is that the types of vulnerabilities could strongly correlate to certain code tokens (distinguishing tokens) in several crucial parts of programs. The distinguishing tokens for each \vt are statistically identified based on their prevalence in the type versus the others. Our results show that the baseline approach enhanced by our component can outperform the state-of-the-art deep pre-trained approaches while retaining very high efficiency. Furthermore, the proposed component could also improve the neural network approaches by up to 92.8\% in macro-average F1.

\end{abstract}


\maketitle

\section{Introduction}
Software vulnerabilities are weaknesses in a software system that could be exploited by attackers. This exploitation can cause substantial damage, especially for the critical systems~\citep{survey_3}.
To reduce manual effort in discovering vulnerabilities, researchers have invested considerable effort in investigating effective approaches for \textit{automated vulnerability detection}, leading to many techniques~\citep{linevd,linevul,ivdetect,vuldeeppeaker,sysevr,velvet,poster,vulsniper,devign,are_we_there,cheng2022path}. 
Recent advances in this field have resulted in a (quite) high accuracy in determining whether an entire given method/file is vulnerable or not~\cite{linevul,vuldeeppeaker,sysevr,velvet,poster}. For example, with \textit{BigVul} benchmark~\cite{bigvul}, Fu \etal~\cite{linevul} propose \textit{LineVul}, which achieves more than 90\% accuracy in vulnerability detection at the function level.
With the support of these state-of-the-art techniques, the next step, security analysis, that developers have to perform is investigating the detected vulnerable functions to determine the actual presence of the vulnerability. 
%
%
However, even having vulnerable code, investigating to fix those functions could still be a non-trivial task~\cite{vul_secret,openai_fixing,transfer_learning_for_fixing}. 
%


\hl{Meanwhile, the knowledge of vulnerability types, such as \textit{buffer overflow} or \textit{memory corruption}, can offer significant insights to developers when debugging the vulnerable code. This information serves as a guide to understanding the principles behind the vulnerability, enabling developers to swiftly pinpoint the exact location of the vulnerability and propose potential approaches to rectify the flawed code}~\cite{mu_vuldeeppeker}. 
\hl{For instance, if a developer identifies a vulnerable function, as shown in Figure}~\ref{fig:BO_example}, \hl{and knows it contains a buffer overflow error, she can immediately focus her investigation on the code statements responsible for writing or copying data into the buffer. This allows her to assess the likelihood of these statements exceeding the buffer's boundaries and overwriting adjacent memory locations. Consequently, she can initiate security analysis by examining the code statements at lines 7, 19, and 30 instead of analyzing the entire function. Once the vulnerable statement (line 30) is identified, she can employ established techniques, such as implementing size checks before writing or correcting the amount of data to be copied, to mitigate the vulnerability. Thus, by determining the vulnerability types after detecting vulnerable code, developers and code auditors can significantly reduce their workload, which is especially beneficial when dealing with large sections of vulnerable code.}

\begin{figure}
    \centering
\begin{lstlisting}[language=C]
static long ec_device_ioctl_xcmd(struct cros_ec_dev *ec, void __user *arg)
{
	long ret;
	struct cros_ec_command u_cmd;
	struct cros_ec_command *s_cmd;

	if (<@\textcolor{red}{copy\_from\_user(\&u\_cmd, arg, sizeof(u\_cmd))}@>)
		return -EFAULT;

	if ((u_cmd.outsize > EC_MAX_MSG_BYTES) ||
	    (u_cmd.insize > EC_MAX_MSG_BYTES))
		return -EINVAL;

	s_cmd = kmalloc(sizeof(*s_cmd) + max(u_cmd.outsize, u_cmd.insize),
			GFP_KERNEL);
	if (!s_cmd)
		return -ENOMEM;

	if (<@\textcolor{red}{copy\_from\_user(s\_cmd, arg, sizeof(\*s\_cmd) + u\_cmd.outsize)}@>) {
		ret = -EFAULT;
		goto exit;
	}

	s_cmd->command += ec->cmd_offset;
	ret = cros_ec_cmd_xfer(ec->ec_dev, s_cmd);
	/* Only copy data to userland if data was received. */
	if (ret < 0)
		goto exit;

	if (<@\textcolor{red}{copy\_to\_user(arg, s\_cmd, sizeof(*s\_cmd) + u\_cmd.insize)}@>)
		ret = -EFAULT;
exit:
	kfree(s_cmd);
	return ret;
}
\end{lstlisting}
        \caption{\hl{A \textit{buffer overflow} in Linux Kernel, CVE-2016-6156}}
        \label{fig:BO_example}
\end{figure}

Despite the importance of \hl{vulnerability type identification (VTI) in debugging vulnerable code after being detected}, the problem has not received the deserved attention. In this work, we make the first step to explore the problem of VTI by using the existing techniques in software engineering (SE) and natural language processing (NLP). Particularly, we model VTI as the multi-label text classification task in NLP~\cite{multilabel_cls,text_multi_label,brm}. \hl{This is reasonable because each vulnerable function $f$ could be considered as a document $d$, and the type set of the vulnerabilities in $f$ could be the label set of $d$}. 

The NLP community recently witnessed a dramatic paradigm shift towards the ``\textit{pre-training + fine-tuning}'' framework. Deep pre-trained models, e.g., BERT~\cite{bert}, induce powerful embeddings that can be rapidly fine-tuned on many downstream NLP problems by adding a task-specific lightweight linear layer on top of the transformer models. BERT-like models (e.g., XLNet~\cite{xlnet} and RoBERTa~\cite{roberta}) have led to state-of-the-art performance on many NLP tasks, such as part-of-speech tagging or text classification.
For the SE, pre-trained models have recently shown to be highly effective in many classification tasks such as bug detection, clone detection, and vulnerability detection~\cite{zhang2019novel,embedding_emse22,embedding_icse19,assessing_codebert, linevd,linevul}. This naturally raises a question: \textit{How are these pre-trained models effective for the VTI task?}

Additionally, we suspect that given a set of all possible vulnerability types $T$ and \hl{a vulnerable function $f$, the types of the vulnerability in $f$}, $S \subseteq T$ could be determined by the appearance of code tokens in $f$. 
For example, \textit{directory traversal} vulnerabilities, which allow attackers to access files/directories stored outside the web root directory, usually contain certain code (sub)tokens such as \texttt{file}, \texttt{base\_path}, or \texttt{directory}. Another example is that \textit{buffer overflow} errors usually associate with \texttt{buffer} (\texttt{buf}) or \texttt{copy} (\texttt{cpy}).
Meanwhile, the importance or relevance of these (sub)tokens in determining the types of vulnerabilities could be captured well by the old-fashioned code representations such as TF-IDF features. Another natural question is: \textit{How does a simple classification model with an old-fashioned code representation such as TF-IDF work for the VTI task compared to pre-trained models?}

In this paper, we conduct experiments to evaluate the performance of the state-of-the-art methods in natural language processing (NLP) and software engineering (SE) for the VTI task. We select two pre-trained models, Word2vec~\cite{word2vec_1} and CodeBERT~\cite{codebert}, which are the representative non-contextual and contextual code embeddings~\cite{ese_empirical}. We pick Word2vec because this model has become one of the most popular pre-trained models for code due to its efficiency~\cite{zhang2019novel,embedding_emse22}. Meanwhile, the reasons for our selection of CodeBERT are the model's reputation and its strong improvements in many SE tasks~\cite{assessing_codebert}. We also compare these state-of-the-art models against a simple multi-label classification model, Binary Relevance (BR), with old-fashion TF-IDF features. With the BR classifier and TF-IDF, this approach is considered as the baseline of multi-label text classification in NLP~\cite{brm} (so-called \base).

%
%
Our experiments on BigVul benchmark~\cite{bigvul} show a \textit{surprising result} that \textit{the advanced pre-trained models just slightly improve the VTI performance of the baseline approach by only less than 7\% in classification accuracy}. Meanwhile, the deep pre-trained models require GPUs and cost up to 40X and 5X more time in training and predicting compared to the baseline approach. These show that for the VTI task, TF-IDF and classical Binary Relevance could capture well the features to distinguish vulnerability types and \textit{efficiently} achieve performance very competitive with the deep pre-trained models in VTI.

In this work, we also introduce a simple technique to improve the VTI performance of \base. Our idea is that certain code (sub)tokens are more likely to appear/not appear in the vulnerabilities of a type than the others. 
%
%
%
%
These \textit{distinguishing tokens} are identified beforehand (before predicting) by statistically analyzing the syntactic code elements crucial for VTI (e.g., \textit{function calls}, \textit{assignments}, or \textit{control structures}) in previously known vulnerable code.
As an \textit{independent component}, these distinguishing tokens are used to refine the predictions produced by \base. 
The intuition is that if a function is predicted to contain a vulnerability \textit{not} of type $t$, yet actually contains the distinguishing tokens of $t$ which are prevalent in the cases of $t$ but not the other types; then the prediction will be refined to the vulnerability of $t$.

Our experimental results show that the combination of \base and our component significantly improves the VTI performance of \base and outperforms the advanced pre-trained VTI models. Meanwhile, the predicting time slightly increases, and identifying distinguishing tokens does not increase the overall training and preparing time. These results indicate that \textit{\base combining with a very lightweight component could improve the VTI performance of advanced pre-trained models while retaining a very high efficiency}. We also show that our technique could effectively improve the other VTI approaches when applied on top of them up to 92.8\% F1-score.

 In brief, this paper makes the following contributions:
\begin{enumerate}
    \item An exploratory study on the performance of both traditional and advanced VTI techniques.
    \item Surprising experimental results showing that a simple model with an old-fashioned feature extraction could achieve a very competitive performance with the state-of-the-art approaches.
    \item A lightweight but effective technique improving the performance of the existing approaches.
\end{enumerate}
The detailed implementation and dataset can be found at: \textit{\url{https://github.com/sonnguyenvnu/VIT-Project}}.

The rest of this paper is organized as follows. Section~\ref {sec:problem} states the problem of vulnerability type identification (VTI) with a benchmark and evaluation metrics for VTI approaches. Then, several VTI approaches modeling the VTI task as the multi-label classification task and our surprising results are introduced in Section~\ref{sec:approaches}. 
After that, Section~\ref{sec:our_approach} introduces the design of our lightweight independent technique and its effectiveness in improving \base as well as the other approaches. 
Some threats to validity are discussed in Section~\ref{sec:threats}.
Section~\ref{sec:related_work} provides the related work. Finally, Section~\ref{sec:conclusion} concludes this paper.
\section{Vulnerability Type Identification}
\label{sec:problem}
Given a finite set of vulnerability types $T$ and a vulnerable function $f$, the vulnerability type identification (VTI) task associates a subset of types $S \subseteq T$ with the function $f$. \hl{A vulnerable function dataset $D$ consists of $N$ vulnerable functions with their vulnerability types $(f_1, S_1)$, $(f_2, S_2)$,... , $(f_N, S_N)$}. 
In this work, \textit{we model the problem of vulnerability type identification as multi-label text classification}~\citep{multilabel_cls}. 
\hl{This is reasonable because each vulnerable function could be considered as a document, and the types of vulnerabilities in the function could be considered as the label/tag set of the document. Note that each function in $D$ is a vulnerable one which could be effectively detected by the existing vulnerability detection techniques}~\cite{ivdetect,linevd,linevul,vuldeeppeaker,sysevr,velvet,poster}.

\subsection{Dataset}
\textit{Data selecting.} To evaluate VTI approaches, we use BigVul, which is one of the largest vulnerability datasets and provides the vulnerability types of each case. The dataset is collected from 348 real-world C/C++ projects on GitHub, such as Chromium, Linux, Android, PHP, OpenSSL, QEMU, and FFmpeg. The dataset includes about 10,900 vulnerable functions in 13 vulnerability types and 44,603 vulnerable lines of code. 
\hl{In this dataset, the types of each vulnerable function are extracted from its corresponding CVE-Details.}

\begin{figure}
    \centering
    \includegraphics[width=1\columnwidth]{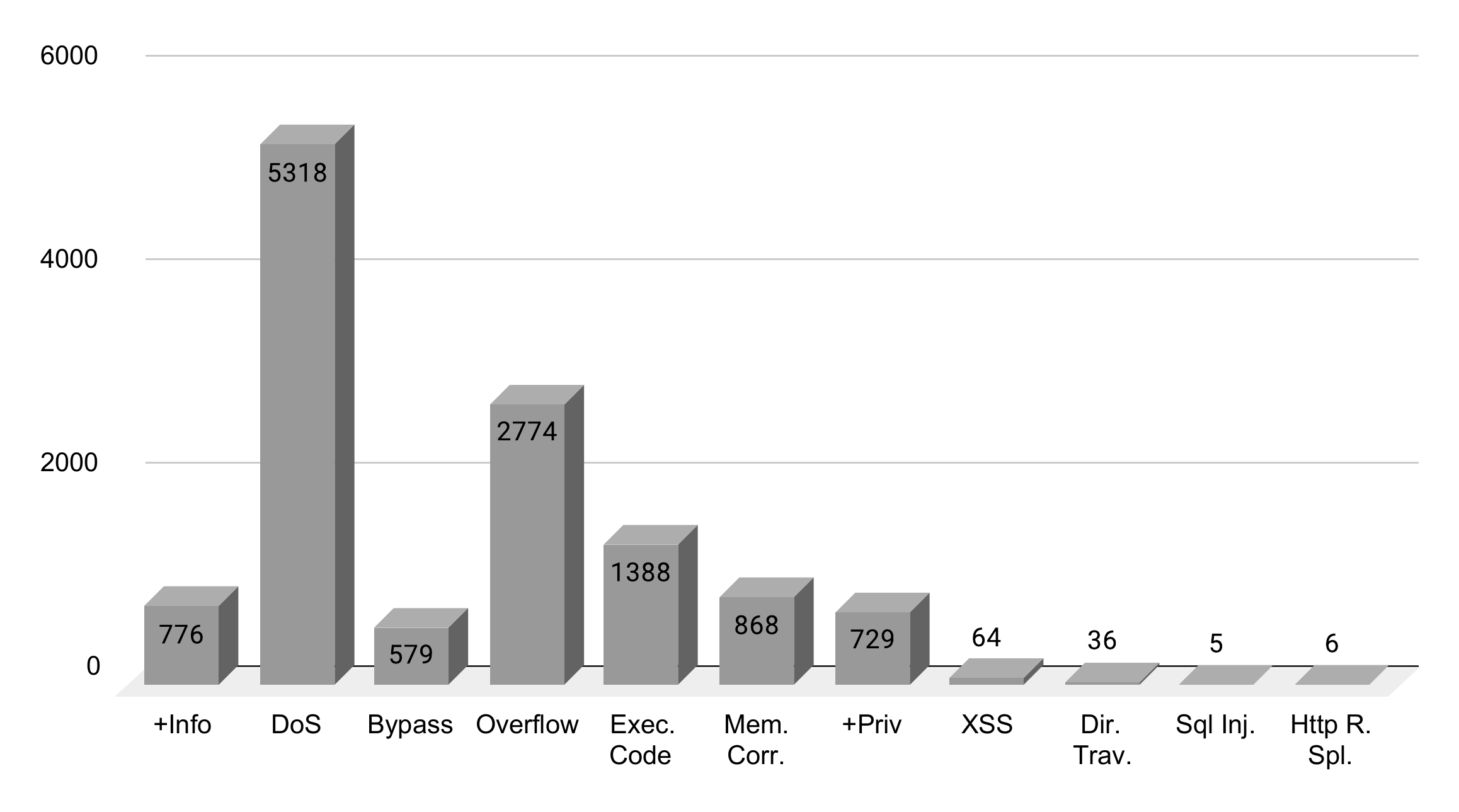}
    \caption{The number of cases by vulnerability types}
    \label{fig:stat_types}
\end{figure}

\begin{figure}
    \centering
    \includegraphics[width=1\columnwidth]{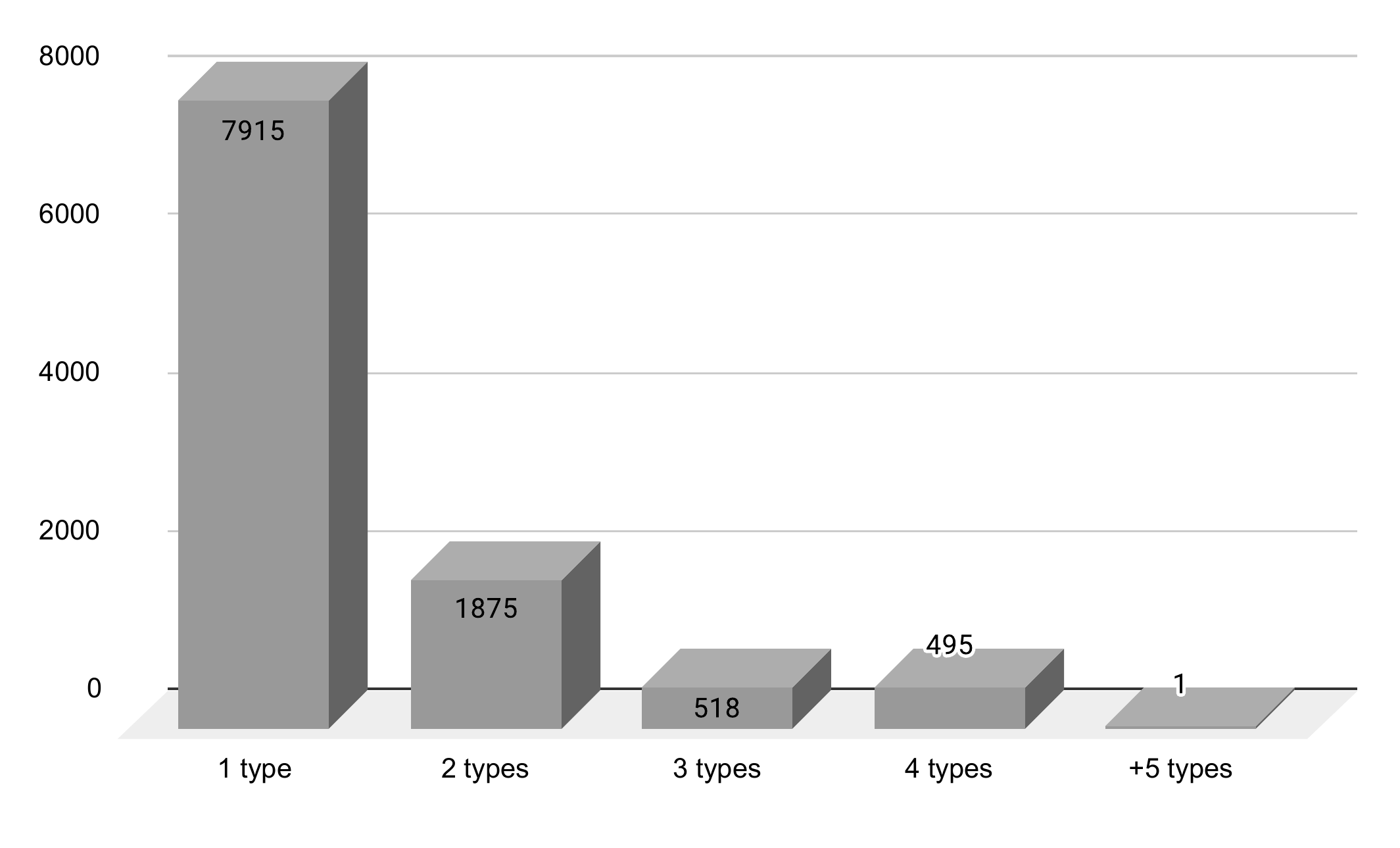}
    \caption{The number of cases by the number of types}
    \label{fig:stat_num_labels}
\end{figure}

Fig.~\ref{fig:stat_types} and Fig.~\ref{fig:stat_num_labels} show the statistical information of BigVul on vulnerability types and the number of types of vulnerability in each function. Particularly, \textit{Denial of Service (DoS)} and \textit{Overflow} are the two most popular types of vulnerability in the dataset. 
This is expected because \textit{Denial of Service (DoS)} and \textit{Overflow} are two of the most frequent vulnerability types~\cite{vul_type_stat}.
Meanwhile, \textit{Sql Injection (Sql Inj.)} and \textit{Http Response Splitting (Http R.Spl)} are very rare in BigVul. As seen in Figure~\ref{fig:stat_num_labels}, the vulnerabilities in most of the functions belong to only one type (about 72\%). \hl{Additionally, there are more than 1K vulnerable functions having more than two types.} 

There are some other public vulnerability datasets, but they are not suitable for being used in our experiments. Particularly, SATE IV Juliet~\cite{juliet} is a synthetic vulnerability dataset that is manually created from known vulnerable patterns. However, Chakraborty \etal~\cite{are_we_there} have demonstrated that this dataset contains only simple vulnerabilities which do not reflect real-world vulnerabilities. 
Meanwhile, Devign~\cite{devign} and Reveal~\cite{are_we_there}, which are constructed from the vulnerabilities in real-world open-source projects, do not provide the vulnerability types for the vulnerable functions. Thus, we do not use these vulnerability datasets in our evaluation experiments.

\textit{Data splitting.} To evaluate VTI methods, we use a random training/validation/test split ratio of 80:10:10, i.e., the whole BigVul is split into 80\% of training data, 10\% of validating/fine-tuning data, and 10\% of testing data.

\subsection{Evaluation Metrics}
To evaluate VTI approaches, we apply the evaluation metrics which are widely used in multi-label classification studies~\citep{text_multi_label}. For each function $f_i$ in the test set, the types of $f_i$ is represented by a vector $Y_i$ such that for $j \in [1,\|T\|]$, $Y_{ij} = 1$ if $T_j \in T$ is one of the vulnerability types of $f$, otherwise $Y_{ij} = 0$. Let denote vector $Z_i$ the vulnerability type prediction for $f_i$ produced by a VTI technique. For a test set containing $N$ cases (functions), the considering evaluation metrics including \textit{exact match ratio}, \textit{hamming score}, \textit{accuracy}, \textit{macro-average}, \textit{micro-average}, \textit{weighted-average}, \textit{sample-average}, are computed as follows.

\textit{Exact match ratio} indicates the percentage of vulnerable functions (cases) that have \textit{all labels} predicted correctly:
$$
\text{\textit{ExactMatchRatio}} = \frac{1}{N}\sum_{i=1}^{N} \mu(Y_i = Z_i)
$$
where $\mu$ returns $1$ if $Y_i$ and $Z_i$ are exactly matched, $\forall j \in [1,\|T\|], Y_{ij} = Z_{ij}$, and 0 otherwise. \textit{Exact match ratio} is the most strict one among the metrics. We also use other less strict metrics for the multi-labels classification task.

\textit{Hamming score} is defined as the proportion of the correctly predicted types to the total number of predicted types and actual types for each case. The overall hamming score is the average across all cases.
$$
\text{\textit{HammingScore}} = \frac{1}{N}\sum_{i=1}^{N} \frac{\|Y_i \cap Z_i\|}{\|Y_i \cup Z_i\|}
$$
Note that $\|Y_i \cap Z_i\| = \sum_{j=1}^{\|T\|} (Y_{ij} \wedge Z_{ij})$ and $\|Y_i \cup Z_i\| = \sum_{j=1}^{\|T\|} (Y_{ij} \vee Z_{ij})$. 
As seen, \textit{Hamming score} only considers label $1$ in $Y_i$ and $Z_i$. We additionally use \textit{accuracy}, which considers matching for both labels $1$ and $0$.

In this work, \textit{accuracy} is calculated as the following formula:
$$
\text{\textit{Accuracy}} = \frac{1}{N} \sum_{i=1}^{N} \frac{1}{\|T\|} \sum_{j=1}^{\|T\|} \mu(Y_{ij} = Z_{ij})
$$

There are several different methods to measure a multi-label classifier by averaging out the types: \textit{micro-averaging}, \textit{macro-average}, \textit{weighted-average}, and \textit{sample-average}. For \textit{micro-averaging}, all true-positive cases (TPs), true-negative cases (TNs), false-positive cases (FPs), and false-negative cases (FNs) for each type are summed up, and then the average is taken. In the micro-averaging method, we sum up the individual TPs, FPs, and FNs of the system for different sets and then apply them. 

$$
Precision^{micro} = \frac{ \sum_{t \in T} TPs(t) }{ \sum_{t \in T} TPs(t) + FPs(t) }
$$

$$
Recall^{micro} = \frac{ \sum_{t \in T} TPs(t) }{ \sum_{t \in T} TPs(t) + FNs(t) }
$$
And the \textit{micro-average} F1-score will be simply the harmonic mean of the above two equations.
$$
F1^{micro} = \frac{2 \times Precision^{micro} \times Recall^{micro}}{Precision^{micro} + Recall^{micro}}
$$

\textit{Macro-average} is straight forward. We just take the average of the precision and recall of the system on different sets.

$$
Precision^{macro} = \frac{ \sum_{t \in T} Precision(t) }{ \|T\| }
$$

$$
Recall^{macro} = \frac{ \sum_{t \in T} Recall(t) }{ \|T\| }
$$

$$
F1^{macro} = \frac{2 \times Precision^{macro} \times Recall^{macro}}{Precision^{macro} + Recall^{macro}}
$$

\textit{Weighted-average} is simply the average of the precision and recall for individual classes weighted by the support of that class. Meanwhile, to compute \textit{Sample average}, precision, recall, and F1-score are computed for each case and then averaged them.
\section{Vulnerability Type Identification as Multi-label Classification}
\label{sec:approaches}
In general, multi-label classification could be addressed by two main approaches: \textit{problem transformation methods} and \textit{adapted methods}. \textit{Problem transformation methods} transform a multi-label problem into multiple binary classification problems. In this fashion, the binary classifier of each label $l \in L$ can be employed to make the classifications, and these are then transformed back into multi-label representations.
Meanwhile, \textit{adapted methods} adapt existing binary classification approaches to tag items with multiple labels without requiring problem transformations.

\subsection{Classical Baseline Approach}
The baseline approach, which is called the binary relevance (BR) method~\citep{brm}, transforms a multi-label problem into one binary classifier for each label. Hence BR independently trains $\|L\|$ binary classifiers $C_1,... , C_{\|L\|}$. Each classifier $C_i$ is responsible for predicting the 0/1 association for each corresponding label $l_i \in L$.
This approach is popular because of its conceptual simplicity, but this method ignores label correlations. Due to this information loss, BR's predicted label sets are likely to contain either too many or too few labels or labels that would never co-occur in practice~\cite{brm}.

For the baseline approach, we use simple TF-IDF (short for Term Frequency – Inverse Document Frequency) features to represent the instances. Basically, every word in the vocabulary set is considered as a feature. Each function is represented as a bag-of-word vector. In this vector, the value of a feature (word) increases proportionally to its count in the function, but it is inversely proportional to the frequency of the word in the corpus. 
The reason for the selection of TF-IDF for the baseline approach is that TF-IDF has been applied as the baseline representation and shown its potential in various software engineering tasks~\citep{embedding_emse22,embedding_icse19} such as code authorship identification~\citep{authorship} or defect prediction~\citep{defect_prediction_paper}. 
Indeed, the existing studies have empirically shown that the use of TF-IDF could outperform the methods using more sophisticated approaches for certain tasks~\cite{embedding_emse22,embedding_icse19}, such as the task of Code Authorship Identification.

\begin{figure}
    \centering
    \includegraphics[width=0.65\columnwidth]{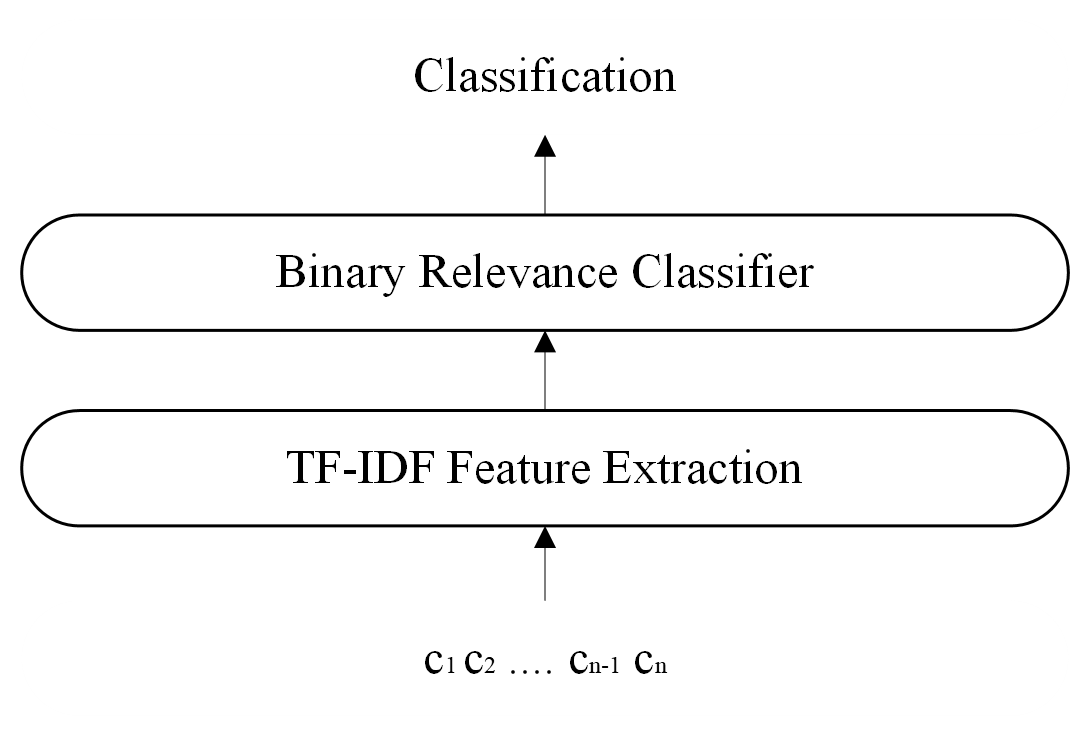}
    \caption{Traditional Baseline Approach (\base) for vulnerability type identification}
    \label{fig:tf-idf}
\end{figure}

In this work, we construct the feature set covering unigrams and bigrams. In order to drop some irrelevant features and reduce the dimensionality problem, we apply the Chi-Square test for feature selection to select the most statistically relevant TF-IDF features, i.e., keep only the features exceeding a certain $p$-value. The reduced vectors are fed to a Binary Relevance Multi-label Classifier with a Gaussian Naive Bayes-based Classifier (Fig.~\ref{fig:tf-idf}).

\subsection{Deep Learning Approaches}
Neural networks (NN) can be directly adapted to support multi-label classification by simply specifying the number of target labels as the number of nodes in the output layer. For example, a task that has three output labels (classes) will require a NN output layer with three nodes in the output layer.
Additionally, till now, the community of SE researchers has paid tremendous efforts to develop powerful code representations for SE classification tasks, such as code authorship identification, code clone detection, source code classification, and software defect prediction. In this work, we investigate the performance of multi-label classification models for VTI adapted from NN approaches using advanced code representations.

With the rapid development of deep learning in SE applications, various code representation techniques have been proposed, which can be categorized into two broad categories: \textit{Non-contextual Embeddings} and \textit{Contextual Embeddings}. Non-contextual embeddings such as Word2vec~\citep{word2vec_1}, GloVe~\citep{glove}, fastText~\citep{fasttext}, Code2vec~\citep{code2vec}, produce \textit{fixed} representations for words in the vocabulary without considering the meanings of words/code tokens in different contexts. Meanwhile, by contextual embeddings such as CodeBERT~\citep{codebert} and CuBERT~\citep{cubert}, the representations of tokens are \textit{adjusted} based on different contexts.

For \textit{Non-contextual Embeddings}, Word2vec has become one of the most popular code embedding techniques for software engineering tasks~\citep{zhang2019novel,embedding_emse22} due to its high efficiency. Word2vec produces a low-dimensional semantic space by using two different model architectures: Skip-gram (i.e., starting from a single word to predict its context) or Continuous Bag-of-Words (i.e., starting from the context to predict a word).
In this work, we use Word2vec with Skip-gram as a representative non-contextual embedding technique.

For \textit{Contextual Embeddings}, we select CodeBERT~\citep{codebert} as this model has achieved strong improvements on many SE tasks, showing their great generalizability~\citep{assessing_codebert, embedding_emse22}. 
CodeBERT shares the same architecture of BERT (Bidirectional Encoder Representation from Transformer)~\citep{bert}, which uses the bidirectional transformer encoder to effectively exploit both the left and right contexts of a target token. Two objectives are designed for BERT-liked models: masked language model and next sentence prediction. In the masked language model, some of the tokens are randomly masked, and the goal is to predict these masked tokens based on their surrounding unmasked context tokens. For next-sentence prediction, the goal is to predict whether a sentence is the next sentence of the current one to capture the relationships between sentences.

To the best of our knowledge, no vulnerability type identification study has been published. 
Note that $\mu$VulDeePecker~\cite{mu_vuldeeppeker} could output vulnerable functions with vulnerability type. However, that approach's goal is to determine if a function is clean or of exactly one vulnerability, this is fundamentally different from the task of vulnerability type identification. The detailed differences between VTI approaches and $\mu$VulDeePecker are discussed in Section~\ref{sec:related_work}.

\begin{figure}
    \centering
    \includegraphics[width=0.65\columnwidth]{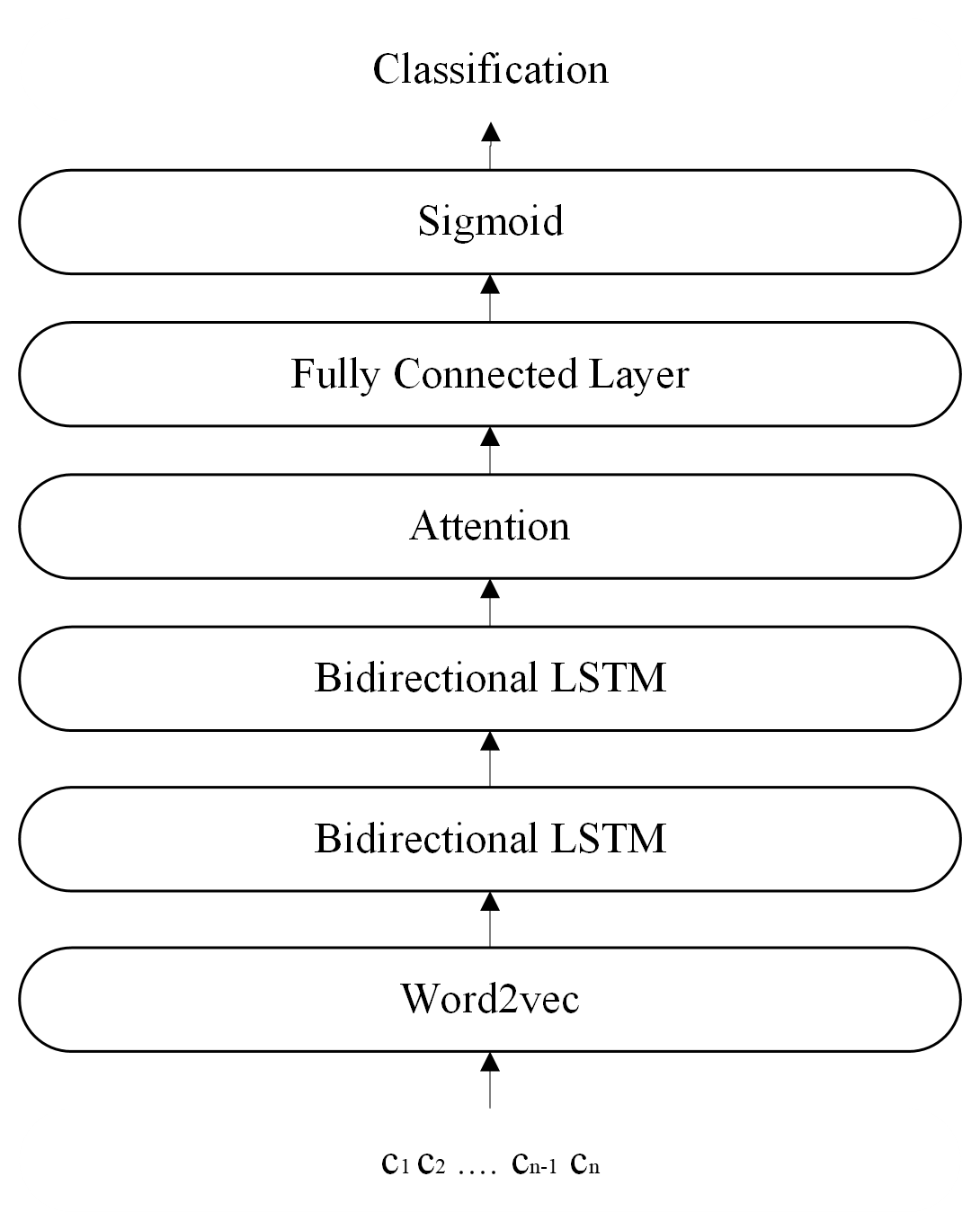}
    \caption{Vulnerability type identification with Word2vec}
    \label{fig:w2v}
\end{figure}

\begin{figure}
    \centering
    \includegraphics[width=0.65\columnwidth]{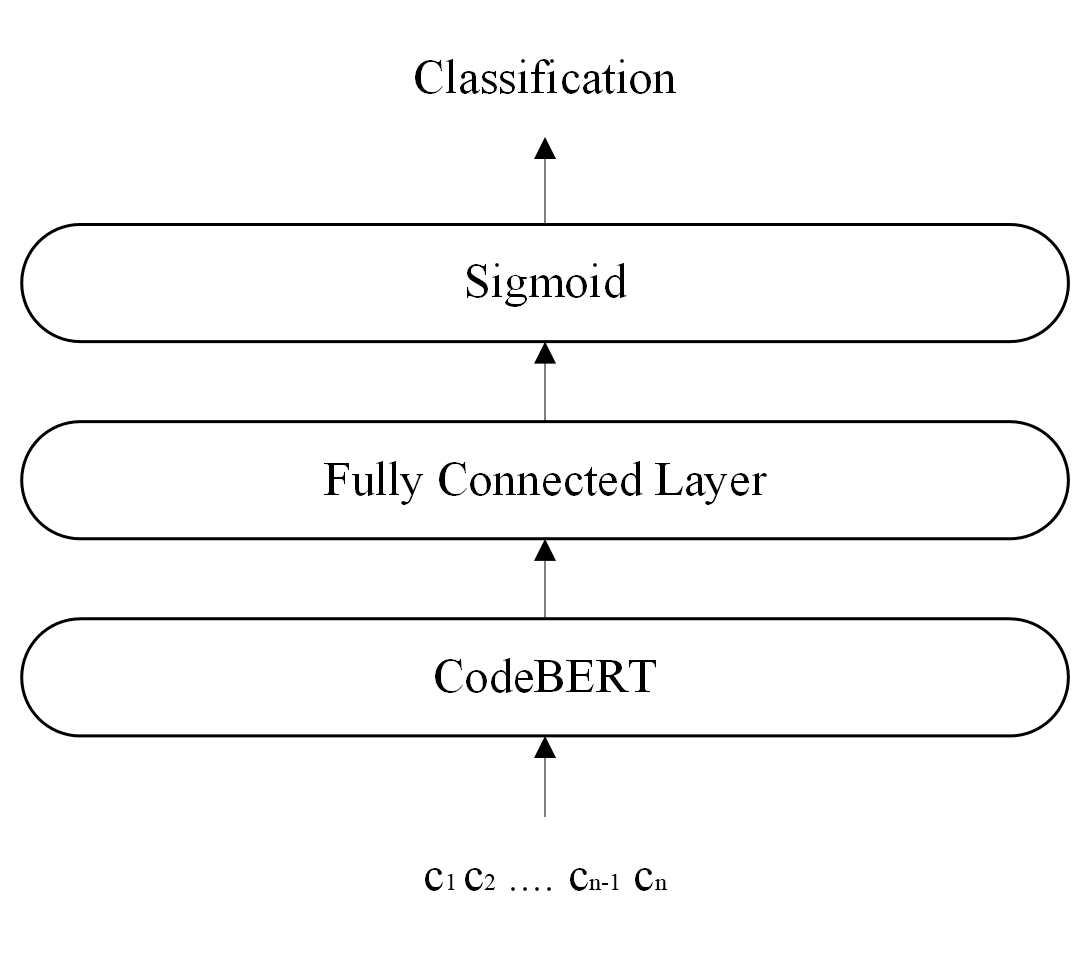}
    \caption{Vulnerability type identification with CodeBERT}
    \label{fig:codebert}
\end{figure}

In this work, for both deep learning models, we follow the typical architectures of classifiers for the general multi-label text classification task instead of designing a complex model.
We evaluate the performance of the multi-label classification model using Word2vec proposed by~\cite{text_multi_label}. They use Word2vec to construct a word embedding layer followed by two Bi-LSTM layers, an attention layer, a fully connected, and the sigmoid activation function (Fig.~\ref{fig:w2v}).
Fig.~\ref{fig:codebert} shows the classification model with CodeBERT proposed by~\cite{assessing_codebert}. In this model, CodeBERT is used as an embedding layer encoding every vulnerable function to a vector. The model also feeds the embedding to a fully connected layer and either the sigmoid function to compute the classification.
For both models, we use binary cross entropy~\cite{text_multi_label} to compute the loss between the classification and the ground truth.

\subsection{Experimental Results}
\label{sec:results}


Table~\ref{tab:type_iden_perf} (\textit{Original Performance} tab) shows the performance of the studied VTI approaches. 
Note that all our experiments were conducted on a workstation with a P100 GPU, dual vCPUs, and 32GB RAM.

\begin{table*}[]
\centering
\caption{Vulnerability type identification performance}
\label{tab:type_iden_perf}
\begin{tabular}{@{}ll|rrr|rrr@{}}
\toprule
\multicolumn{2}{l}{} & \multicolumn{3}{c|}{\textit{Original Performance}}      & \multicolumn{3}{c}{\textit{After Enhancement}}       \\ \cmidrule(l){3-8} 
\multicolumn{2}{l}{}                         & \base    & Word2vec  & CodeBERT  & \base    & Word2vec  & CodeBERT  \\ \cmidrule(l){1-8} 
\multirow{3}{*}{\textit{micro-avg}}        & Prec.    & 0.71    & 0.55      & 0.73      & 0.72    & 0.60      & 0.74      \\
                                  & Rec.     & 0.69    & 0.43      & 0.73      & 0.78    & 0.56      & 0.79      \\
                                  & F1       & 0.70    & 0.48      & 0.73      & 0.75    & 0.58      & 0.77      \\
                                  \midrule
\multirow{3}{*}{\textit{macro-avg}}        & Prec.    & 0.65    & 0.41      & 0.67      & 0.69    & 0.70      & 0.70      \\
                                  & Rec.     & 0.60    & 0.24      & 0.65      & 0.75    & 0.48      & 0.76      \\
                                  & F1       & 0.62    & 0.28      & 0.65      & 0.71    & 0.54      & 0.72      \\
                                  \midrule
\multirow{3}{*}{\textit{weighted-avg}}     & Prec.    & 0.71    & 0.50      & 0.73      & 0.73    & 0.61      & 0.74      \\
                                  & Rec.     & 0.69    & 0.43      & 0.73      & 0.78    & 0.56      & 0.79      \\
                                  & F1       & 0.70    & 0.47      & 0.73      & 0.75    & 0.55      & 0.76      \\
                                  \midrule
\multirow{3}{*}{\textit{sample-avg}}       & Prec.    & 0.65    & 0.47      & 0.71      & 0.73    & 0.57      & 0.75      \\
                                  & Rec.     & 0.68    & 0.42      & 0.72      & 0.80    & 0.60      & 0.81      \\
                                  & F1       & 0.65    & 0.42      & 0.70      & 0.74    & 0.55      & 0.76      \\
                                  \midrule
\multicolumn{2}{l|}{\textit{ExactMatchRatio}}          & 0.54    & 0.30      & 0.57      & 0.60    & 0.32      & 0.57      \\
\multicolumn{2}{l|}{\textit{HammingScore}}            & 0.62    & 0.39      & 0.66      & 0.69    & 0.49      & 0.71      \\
\multicolumn{2}{l|}{\textit{Accuracy}}                 & 0.89    & 0.84      & 0.90      & 0.91    & 0.86      & 0.91      \\
\midrule
\multicolumn{2}{l|}{\textit{Training time (ms)}}            & 148,824 & 2,610,032 & 4,938,020 & 148,851 & 2,610,035 & 4,938,035 \\
\multicolumn{2}{l|}{\textit{Predict. time (ms)}}            & 4,320   & 12,349    & 22,282    & 5,243   & 13,841    & 24,028    \\
\midrule
\multicolumn{2}{l|}{\textit{GPU required}}             & No      & Yes       & Yes       & No      & Yes       & Yes       \\ \bottomrule
\end{tabular}
\end{table*} 

As seen in \textit{Original Performance} tab of Table~\ref{tab:type_iden_perf}, VTI with CodeBERT achieved the best performance among the three approaches. 
However, the performance of this deep pre-trained model is slightly better than that of \base, about 4\%-7.5\% in F1-scores for \textit{macro average}, \textit{micro average}, \textit{weight average}, and \textit{sample average}. Meanwhile, this advanced method improves 5.5\%, 6.4\%, and only 1.1\% in \textit{exact match ratio}, \textit{hamming score}, and \textit{accuracy}, respectively. \base even significantly outperforms the VTI approach with Word2vec in all the considering metrics. Especially, the micro-average of \base doubles the corresponding figure of Word2vec. 

Table~\ref{tab:perf_type} shows the detailed classification performance (F1-score) of the three approaches for each vulnerability type. \base achieves F1-scores comparable with those of the CodeBERT-based approach, even slightly better for \textit{Memory Corruption} and \textit{Others}. Additionally, \base significantly improves the performance of the Word2vec-based method for all types. 
These results empirically demonstrate that \textit{\base with a traditional multi-label classification model and the old-fashioned TF-IDF could achieve a very competitive performance compared to the advanced VTI approaches}.

Especially, \base is much more efficient than the other approaches. Specially, \base is trained \textbf{20X} and \textbf{40X faster} than the VTI approaches based on Word2vec and CodeBERT. Moreover, \base is also much faster in predicting vulnerability type of vulnerable functions, 3X and 5X faster compared to the Word2vec-based and CodeBERT-based models. Notably, unlike the neural network approaches, \base can be trained and tested \textbf{without GPU}, while the Word2vec-based and CodeBERT-based models require GPU and consume much computational resource. This advantage enables a flexible deployment for \base on developers' machines which might not be very powerful with limited resources while retaining a competitive VIT performance.


\begin{table}[]
\centering
\caption{Performance in F1 score by types}
\label{tab:perf_type}
\begin{tabular}{l|r|r|r}
\toprule
\textit{Types}  & \base     & Word2vec & CodeBERT \\ \midrule
Information Gain           & 0.41      & 0.00         & \textbf{0.47}         \\ \midrule
Privilege Gain           & 0.63      & 0.13         & \textbf{0.67}         \\ \midrule
Bypass          & 0.42      & 0.00         & \textbf{0.48}         \\ \midrule
DoS             & 0.81      & 0.68         & \textbf{0.83}         \\ \midrule
Execution Code      & 0.64      & 0.34         & \textbf{0.66}         \\ \midrule
Memory Corruption      & \textbf{0.70}      & 0.36         & 0.68         \\ \midrule
Overflow        & 0.67      & 0.37         & \textbf{0.75}         \\ \midrule
Others          & \textbf{0.69}      & 0.34         & 0.68         \\ \bottomrule
\end{tabular}
\end{table}


\section{A Light-weight Method to Improve Vulnerability Type Identification}
\label{sec:our_approach}

As shown in Section~\ref{sec:results}, \base is \textit{very efficient} because of the adoption of the simple code representation and classification model. However, the appearance of code tokens in all the parts of vulnerable functions is considered equally important. 
Meanwhile, we observe that there are certain code syntactic elements, such as \textit{function calls}, \textit{assignments}, or \textit{control structures}, which could be more crucial than the others in determining vulnerability types. 
Thus, instead of considering code tokens in all code syntactic elements equally important, focusing on those crucial code syntactic elements in vulnerable functions could improve the accuracy of VTI predictions.
We propose an independent prediction-refining component to enhance the VIT performance of \base while retaining its overall efficiency.

\subsection{Design}
%
%
%
%
%
%
%
%
%


\begin{definition}{\textit{\textbf{(Syntactic Code Element).}}}
A syntactic code element is a syntactical part of programs defined by the programming language in use. 
\end{definition}

In this work, we use Joern~\cite{joern}, which is widely applied in the existing studies~\cite{vuldeeppeaker,sysevr,ivdetect,linevd,velvet}, to analyze vulnerable functions and extract their syntactic code elements.
In BigVul, about 95\% of vulnerable statements are/contain either \textit{function calls} (78\%), \textit{assignments} (44\%), \textit{control structures} (38\%), or \textit{return statements} (19\%). Instead of considering all code elements equally important in VTI, focusing on those crucial kinds of elements in vulnerable functions could improve VTI performance.
Based on the observation, we design a lightweight technique that can be applied as an independent step to improve \base's performance. 
Our idea is that the code tokens, which can be used to distinguish each vulnerability type from the others (so-called, \textit{distinguishing tokens}), are identified by statistically analyzing the training set on the selective syntactic code elements which are critical for VTI such as \textit{function calls}, \textit{assignments}, \textit{control structures}, and \textit{return statements}. 
For example, \textit{Buffer Overflow} vulnerabilities usually cause by the \textit{assignments} (e.g., assigning too large index) or \textit{function calls} (e.g., copying data larger a buffer's capacity).
These distinguishing tokens are used to refine the predictions produced by \base. Particularly, a function $f$ should or should not be of a type $t$ if $f$ has the distinguishing tokens of $t$.

\begin{definition}{\textit{\textbf{(Distinguishing Token).}}}
For a type $t \in T$, a distinguishing token in a syntactic code element is a code token which is more/less \textit{prevalent} in the syntactic code elements of the functions having vulnerabilities of type $t$ than in any other types.
\end{definition}
Among $D_t$, which is the collection of cases of type $t$, the prevalence of a token $c$ regarding a syntactic code element $e$, ($prev(c, e, D_t)$), is reflected via the ratio of the syntactic code elements $e$ of cases containing $c$ in $D_t$. Formally, $prev(c, e, D_t) = \frac{count(c, e, D_t)}{\|D_t\|}$, where $count(c, e, D_t)$ is the number of cases where the elements $e$ contain $c$. Token $c$ is a \textit{positive distinguishing tokens} of $t$, $c \in \Delta^+(t, e)$, if $c$ is more prevalent in the cases of type $t$ than any other types: 
$$
dis^+(c, e, t, D) = \frac{prev(c, e, D_t)}{\max_{t' \in T \setminus \{t\}} prev(c, e, D_{t'})} > 1
$$
When $\max_{t' \in T \setminus \{t\}} prev(c, e, D_{t'}) = 0$, then $dis^+(c, e, t, D)$ is infinity. In that case, $c$ appears in $D_t$ only, not in the others.

Meanwhile, regarding a syntactic code element $e$, there are certain tokens which are more prevalent in the cases of the other types rather than $t$. Such code tokens are considered as \textit{negative distinguishing tokens} of $t$ regarding $e$, $c \in \Delta^-(t, e)$. Formally, $c \in \Delta^-(t, e)$ if $c$ satisfies the following condition:
%
$$
dis^-(c, e, t, D) = \frac{\min_{t' \in T \setminus \{t\}} prev(c, e, D_{t'})}{prev(c, e, D_t)} > 1
$$
%
%
%
%
If $c$ is never in any case of type $t$, then $prev(c, e, D_t) = 0$ and $dis^-(c, e, t, D) = \infty$.
Intuitively, regarding a syntactic code element, \textit{when a case has a positive distinguishing token of type $t$, then the vulnerable function is likely to have a vulnerability of $t$} (\rp). Similarly, \textit{if the case has a negative distinguishing token of type $t$, the vulnerable function is likely to be a case of any types other than $t$} (\rn).

A vulnerable function $f$ might satisfy neither \rp nor \rn. The prediction for $t$ has to rely on the prediction of \base. Thus, we apply \rp and \rn to design a technique which can be used as \textit{\textbf{an independent component}} combined with the VTI model. Particularly, for the prediction $Z$ of $f$ produced by \base, we apply \rp and \rn to refine $Z$ to produce the final prediction $Z'$. Particularly, for a type $t \in T$, $Z'[t]$, which is the syntactic code element corresponding $t$ in $Z'$, is refined based on that of $Z$ (i.e., $Z[t]$):
\begin{itemize}
    \item If a syntactic code element $e$ of $f$ contains $c \in \Delta^+(t, e)$, yet $Z[t] = 0$, then $Z'[t] = 1$.
    \item If a syntactic code element $e'$ of $f$ contains $c' \in \Delta^-(t, e)$, yet $Z[t] = 1$, then $Z'[t] = 0$.
    \item Otherwise, $Z'[t] = Z[t]$.
\end{itemize}

In fact, applying the observation, that certain syntactic code elements are more important than others in VTI, as a lightweight independent component could expand the applicability of our technique.
Indeed, the refining step using our technique could be applied as an independent component of any VTI approach. 
%
%
We will show the performance of the other approaches when combined with our technique in Sec.~\ref{sec:our_results}.

\subsection{VIT Performance Improvement}
\label{sec:our_results}

\textbf{\textit{Improving \base's VTI performance}}. To evaluate the effectiveness of our method in improving VIT performance, we apply our method as the post-processing step of \base (so-called \ebase). Note that all our experiments were conducted on a workstation with a P100 GPU, dual vCPUs, and 32GB RAM.

After enhancement, 308 predictions in 1,055 cases are affected with an accuracy rate of 84\%. In other words, there are 260/308 predictions are accurately corrected. Table~\ref{tab:type_iden_perf} (tab \textit{After Enhancement}) shows the VIT performance of \base with the post-processing step. 
Compared to \base (tab \textit{Original Performance}), after applying the post-processing step, the macro-average precision increases by 6.1\%, while the macro-average recall is significantly improved by 25\%. This means that for a \vt $t$, \ebase not only identified much more the cases of $t$ (higher recall) but also is more precise in identifying $t$ (higher precision). 
Indeed, as shown in Table~\ref{tab:improv_base}, the precision of 5/8 types and recall of all types are improved. Especially, the precision for \textit{Information Gain} is improved by 37\%, while \ebase doubles the recall of \base for this type.

For the micro-average in Table~\ref{tab:type_iden_perf}, the precision and recall are slightly improved by 1.4\% and 13.0\%. The reason is that for certain types, such as \textit{DoS} or \textit{Overflow}, with more cases than the other types, \base already learns better in identifying these types. Thus, \ebase did not improve \base much for those types. Consequently, micro-average metrics, which are calculated based on individual TP, FP, and FN, were not improved much. This reason explains the slight improvement in the weighted-average. 
Meanwhile, for samples-average, the improvements in the precision and recall are more significant, 12\% and 17\%, respectively. This demonstrates the effectiveness of the \ebase in improving the individual predictions of \base.

\begin{table*}[]
\centering
\caption{VTI performance of \base and \ebase by types}
\label{tab:improv_base}
\begin{tabular}{l|rrr|rrr}
\toprule
\multirow{2}{*}[-2pt]{\textit{Types}}& \multicolumn{3}{c|}{\textit{\base}}      & \multicolumn{3}{c}{\textit{\ebase}}       \\ \cmidrule(lr){2-7}
 & \multicolumn{1}{r|}{\textit{Precision}} & \multicolumn{1}{r|}{\textit{Recall}} & \textit{F1} & \multicolumn{1}{r|}{\textit{Precision}} & \multicolumn{1}{r|}{\textit{Recall}} & \textit{F1}     \\ \midrule

Information Gain           & \multicolumn{1}{r|}{0.41}          & \multicolumn{1}{r|}{0.42}       & 0.41    & \multicolumn{1}{r|}{0.56}                 & \multicolumn{1}{r|}{0.83}             & 0.67    \\ \midrule
Privilege Gain           & \multicolumn{1}{r|}{0.75}          & \multicolumn{1}{r|}{0.55}       & 0.63    & \multicolumn{1}{r|}{0.80}                 & \multicolumn{1}{r|}{0.74}             & 0.77    \\ \midrule
Bypass          & \multicolumn{1}{r|}{0.43}          & \multicolumn{1}{r|}{0.42}       & 0.42    & \multicolumn{1}{r|}{0.55}                 & \multicolumn{1}{r|}{0.75}             & 0.64    \\ \midrule
DoS             & \multicolumn{1}{r|}{0.81}          & \multicolumn{1}{r|}{0.80}       & 0.81    & \multicolumn{1}{r|}{0.81}                 & \multicolumn{1}{r|}{0.89}             & 0.84    \\ \midrule
Execution Code      & \multicolumn{1}{r|}{0.68}          & \multicolumn{1}{r|}{0.61}       & 0.64    & \multicolumn{1}{r|}{0.70}                 & \multicolumn{1}{r|}{0.66}             & 0.68    \\ \midrule
Memory Corruption      & \multicolumn{1}{r|}{0.82}          & \multicolumn{1}{r|}{0.62}       & 0.70    & \multicolumn{1}{r|}{0.82}                 & \multicolumn{1}{r|}{0.63}             & 0.71    \\ \midrule
Overflow        & \multicolumn{1}{r|}{0.65}          & \multicolumn{1}{r|}{0.69}       & 0.67    & \multicolumn{1}{r|}{0.66}                 & \multicolumn{1}{r|}{0.74}             & 0.70    \\ \midrule
Others          & \multicolumn{1}{r|}{0.66}          & \multicolumn{1}{r|}{0.72}       & 0.69    & \multicolumn{1}{r|}{0.66}                 & \multicolumn{1}{r|}{0.72}             & 0.69 \\
\bottomrule 

\end{tabular}
\end{table*}

The improvements in these above metrics result in the increases of all \textit{exact match ratio}, \textit{hamming score}, and \textit{accuracy} (Table~\ref{tab:type_iden_perf}).
%
As seen, \ebase can give 3/5 fully correct predictions. Meanwhile, more than 2/3 predicted types (\textit{hamming score} of 69\%) are accurately given by \ebase. Compared to \base, both the \textit{exactly match ratio} and \textit{hamming score} of \ebase are more than 10\% better. However, the improvement in \textit{accuracy} is only 2.2\%. This is because the cases whose \vt set is small are very popular, and models tend to predict very few types. Meanwhile, \textit{accuracy} considers both labels 0 and 1 in prediction vectors and ground-truth vectors. Thus, the accuracy of \base in each case is already high. This leads to the low improvement by \ebase in \textit{accuracy}.

\textbf{\textit{Compared to CodeBERT-based approach}}. As seen in Table~\ref{tab:type_iden_perf}, \ebase achieved better performance in all the metrics (except micro-average precision) compared to the CodeBERT-based approach. 
For micro-average, although the precision of \ebase is slightly lower than that of CodeBERT (0.72 vs. 0.73), the improvement of \ebase in the recall is more significant (0.78 vs. 0.73). This results in an improvement in micro-average F1. Analyzing the cases where CodeBERT-based can perform well while \ebase did not, we found that these cases have quite complex logic and belong to multiple types which have a causal relationship. For example, \textit{DoS} vulnerabilities could be caused by \textit{Overflow} ones. Meanwhile, \ebase does not consider the relationship between types. Thus, although they have certain distinguishing tokens of a type, \ebase might fail to infer that a vulnerable code also has other types.

In our implementation, the preparing (prep.) step to extract distinguishing tokens and the model training step are performed in parallel. Thus, the total time for model training and preparing is still 2.48 minutes. Meanwhile, the predicting time slightly increases from 4.0 to 5.2 seconds. These time costs are much more efficient than those of CodeBERT.

\begin{table}[]
\centering
\caption{The effects of the post-pocessing step on the VTI approaches}
\label{tab:effects}

\begin{tabular}{@{}l|rrr@{}}
\toprule
                            & \base     & Word2vec & CodeBERT       \\ \midrule
No. of affected predictions & 308       & 428      & 237            \\ \midrule
Accuracy rate             & 0.74      & 0.80     & 0.72           \\ \bottomrule
\end{tabular}%

\end{table}

\textit{\textbf{Overall},} we can conclude that \textit{a simple model combined with a lightweight component could achieve better VTI performance and be much more efficient than advanced deep pre-trained approaches}. This could be very meaningful for users who want to achieve both high effectiveness and efficiency in VTI.

\textbf{\textit{Effectiveness in improving approaches}}. Inspired by the success of our method in improving the performance of \base, we apply the technique to enhance the other approaches. 
Table~\ref{tab:type_iden_perf} (tab \textit{After Enhancement}) shows the VTI performance of Word2vec and CodeBERT approaches after applying our method as a post-processing step. 
Compared to their performance before enhancement (Table~\ref{tab:type_iden_perf}), the performance of all these approaches is improved. The improvements for CodeBERT are minor but still visible by up to 10\% in macro-average F1. 
The numbers of affected predictions and accuracy rates for Word2vec and CodeBERT are shown in Table~\ref{tab:effects}. As seen, the effect of the post-processing step on the Word2vec-based approach is more significant, with a higher accuracy rate compared to that of the CodeBERT-based method.
The reason could be that the advanced deep pre-trained CodeBERT can capture well certain degrees of our rules. Thus, our method was not very effective in improving the CodeBERT-based VIT approach. 
However, the improvements in all the micro-average, macro-average, weighted-average, and sample-average metrics for Word2vec are significant. Among these metrics, macro-average F1 increases by 92.8\% after applying our post-processing step. These results demonstrate that \textit{our approach is very effective in improving the VTI performance of not only \base, but also the others.}


\

\section{Threats to Validity}
\label{sec:threats}

The main threats to the validity of our work consist of internal, construct, and external threats.

\textbf{Threats to internal validity} include the influence of the method used to identify the code elements (e.g., function calls, assignments, or control structures). To reduce this threat, we use Joern~\cite{joern} code analyzer, which is widely used in existing studies~\cite{velvet,vuldeeppeaker,ivdetect,linevd}. 

\textbf{Threats to construct validity} relate to the suitability of our evaluation procedure. We used \textit{exact match ratio}, \textit{hamming score}, \textit{accuracy}, \textit{macro-average}, \textit{micro-average}, \textit{weighted-average}, and \textit{sample-average}. They are the classical evaluation measures for multi-label classification~\cite{multilabel_cls}. 
%

\textbf{Threats to external validity} mainly lie in the selection of multi-label classification models used in our experiments. To mitigate this threat, we select the representative models which are well-known for NLP and SE tasks. \base is considered as the baseline approach for general multi-label text classification, while all Word2vec, Glove, and CodeBERT are reputed and shown to be effective in many SE tasks. The dataset used in our experiments might not be representative or not very high-quality. To reduce this threat, we used the largest public dataset~\cite{bigvul}, which is collected from a large number of real-world projects and widely used in existing vulnerability detection studies~\cite{linevd, linevul, ivdetect}. Additionally, our data has only C/C++ code. Thus, we cannot claim that similar results would have been observed in other programming languages. Further studies are needed to validate and generalize our findings to other languages. 

\section{Related Work}
\label{sec:related_work}

\textbf{Vulnerability/Bug Detection and Prevention}. Various methods have been proposed to determine if a code component (component, file, function/method, or statement/line) is vulnerable. The rule-based techniques apply static analyzers and leverage seen vulnerability patterns, such as FlawFinder~\cite{FlawFinder} or Coverity~\cite{Coverity}. Recently, several deep-learning based approaches have been introduced~\cite{survey_papers, poster, vuldeeppeaker, vulsniper, devign, are_we_there,mvd,issta_22}. 
VulDeePecker~\cite{vuldeeppeaker} and SySeVR~\cite{sysevr} introduce tools to detect \textit{slice-level} vulnerabilities, which are more fine-grained.
IVDetect~\cite{ivdetect}, which is a graph-based neural network model, is proposed to detect vulnerabilities at the function level and use a model interpreter to identify vulnerable statements in the detected suspicious functions. LineVul~\cite{linevul} and LineVD~\cite{linevd} apply CodeBERT in their own way and have been shown that they are more effective than IVDetect in detecting vulnerable functions and lines/statements. VelVet~\cite{velvet} builds graph-based models to detect vulnerable statements. Our work could complement well with the existing automated vulnerability detection approaches. Particularly, the type identification step could be applied after developers use a vulnerability detection method to quickly interpret and fix the vulnerable functions.
Our work could also be applied to determine the vulnerability types of a vulnerable code component before applying one or more vulnerability detection approaches specialized for certain vulnerability types, such as the approach to memory-related vulnerabilities~\cite{mvd}.

\hl{Our work might related VUDENC}~\cite{VUDENC} \hl{Wartschinski \textit{et al.} and $\mu$VulDeePecker}~\cite{mu_vuldeeppeker} \hl{by Zou \textit{et al.} which focus on vulnerability detection and can implicitly indicate the type of vulnerability. In other words, given a piece of code, these techniques determine if the code is benign or belongs to one/some vulnerability types. VUDENC}~\cite{VUDENC} \hl{uses separate LSTM binary classification models to determine if a piece of code is neutral/benign or belongs to some vulnerability types. 
In this paper, we focus on the specific task of vulnerability type identification integrated after vulnerability detection to provide developers with the type of vulnerabilities effectively identified by the existing vulnerability detection techniques}~\cite{vuldeeppeaker,sysevr,ivdetect,linevul,linevd} \hl{which determine if the given code is benign or vulnerable. }
Meanwhile, the goal of $\mu$VulDeePecker~\cite{mu_vuldeeppeker} is to determine if a function is clean (not vulnerable) or of exactly one vulnerability. However, our work differs from $\mu$VulDeePecker in three fundamental aspects. First, $\mu$VulDeePecker is a vulnerability detection approach which is designed to decide if a function is vulnerable. In other words, the input of that approach is a function that has not known whether it is vulnerable or not. Meanwhile, our work is designed to apply as a step after detecting vulnerable functions. Therefore, the input function of our work is assumed to be vulnerable. Moreover, $\mu$VulDeePecker assumes that a vulnerable function has only one type. Thus, the multi-class classification is considered in $\mu$VulDeePecker. Meanwhile, a function could belong to multiple types. Hence, in our work, the VTI problem is modeled as the multi-label classification task. Finally, $\mu$VulDeePecker is designed for the function call vulnerabilities, while our work has no limit to the kind of vulnerabilities.


\textbf{Learning-based approaches for SE tasks}. Several studies have been proposed for specific SE tasks, including code suggestion/completion~\cite{icse20, naturalness, autosc, apsec21,arist}, program synthesis~\cite{gvero2015synthesizing}, pull request description generation~\cite{hu2018deep,liu2019automatic}, code summarization~\cite{iyer2016summarizing,mastropaolo2021studying,wan2018improving}, code clones~\cite{li2017cclearner}, fuzz testing\cite{godefroid2017learn}, code-text translation~\cite{ase22}, and program repair~\cite{jiang2021cure,ding2020patching}.
Recently, several learning techniques have been proposed to learn representing source code for specific SE applications~\cite{oppsla19, code2vec} or general SE tasks~\cite{codebert,cubert,infercode}.
\section{Conclusion}
\label{sec:conclusion}
In this work, we investigate the problem of vulnerability type identification (VTI) after vulnerability detection. The problem is modeled as the multi-label classification task. Particularly, each detected vulnerable function is considered as a document, and the set of vulnerability types of the function could be considered as the label set of the corresponding document.
This NLP task has been effectively addressed by pre-training, then fine-tuning the framework with deep pre-trained embedding models. 
The existing studies show that the deep pre-trained embedding models specialized for code have also shown their effectiveness for many classification tasks in software engineering.
In this paper, we experimentally evaluate the performance of the well-known and advanced pre-trained models for VTI on a large set of vulnerabilities in various types. Surprisingly, their performance is not much better than the VTI performance of the traditional baseline classification model with an old-fashioned bag-of-word TF-IDF. Meanwhile, these neural network approaches cost much more time and require GPU. We also introduce a lightweight independent component to enhance the predictions of the baseline approach. Our idea is that the types of vulnerabilities could strongly correlate to certain code tokens (distinguishing tokens) in several crucial parts of programs. The distinguishing tokens for a \vt are statistically identified based on their prevalence in the type versus the others. Our results show that the baseline approach enhanced by our component can outperform the state-of-the-art deep pre-trained methods while retaining very high efficiency. Furthermore, the proposed technique could also improve the neural network approaches by up to 92.8\% in macro-average F1.

\printcredits
\bibliographystyle{elsarticle-num}
\bibliography{main}

\end{document}